\documentclass[final,times]{elsarticle}
\usepackage{graphicx,epsfig,amsmath}
\biboptions{sort&compress}

\newcommand{\comment}[1]{}

\newcommand{\bes}{\mbox{$^7$Be }}

\journal{Nuclear Physics A}

\begin{document}

\begin{frontmatter}

\title{Constraining the $^7$Be($p,\gamma$)$^8$B $S$-factor with the new precise $^7$Be solar neutrino flux from Borexino}

\author[hzdr,tud]{M. P. Tak{\'a}cs}
\author[hzdr]{D. Bemmerer \corref{mycorrespondingauthor}}
\cortext[mycorrespondingauthor]{Corresponding author}
\ead{d.bemmerer@hzdr.de}
\author[hzdr]{A. R. Junghans}
\author[tud]{K. Zuber}

\address[hzdr]{Helmholtz-Zentrum Dresden--Rossendorf (HZDR), 01328 Dresden, Germany}
\address[tud]{Technische Universit\"at Dresden,  01069 Dresden, Germany}

\begin{abstract}
Among the solar fusion reactions, the rate of the $^7$Be($p,\gamma$)$^8$B reaction is one of the most difficult to determine rates. In a number of previous experiments, its astrophysical \textit{S}-factor has been measured at $E$ = 0.1-2.5 MeV center-of-mass energy. However, no experimental data is  available below 0.1 MeV. Thus, an extrapolation to solar energies is necessary, resulting in significant uncertainty for the extrapolated $S$-factor. On the other hand, the measured solar neutrino fluxes are now very precise.
Therefore, the problem of the \textit{S}-factor determination is turned around here: Using the measured $^7$Be and $^8$B neutrino fluxes and the Standard Solar Model, the $^7$Be($p,\gamma$)$^8$B astrophysical \textit{S}-factor is determined at the solar Gamow peak. In addition, the $^3$He($\alpha$,$\gamma$)$^7$Be \textit{S}-factor is redetermined with a similar method.
\end{abstract}


\end{frontmatter}

%============================================================================

\section{Motivation}

In the center of the Sun, the competition of the $^7$Be($p,\gamma$)$^8$B reaction with the electron-capture decay of $^7$Be controls the production of radioactive $^8$B in the so-called third proton-proton chain (pp-III). $^8$B, in turn, gives rise to high-energy neutrinos from its $\beta^+$ decay  \cite{Adelberger11-RMP}. Currently, the $^7$Be($p,\gamma$)$^8$B $S$-factor has the highest uncertainty (7.5\% in the solar Gamow peak) of all solar fusion reactions except for the $^3$He(p,e$^+\nu_e$)$^4$He reaction controlling the hitherto unmeasured ''hep'' neutrino flux \cite{Adelberger11-RMP}. Thus, there is a strong motivation to improve the knowledge of the $^7$Be($p,\gamma$)$^8$B $S$-factor, here called $S_{17}(E)$, at solar energies, and this is the aim of the present work.

\begin{figure}[b]
\centering
\includegraphics[width=\columnwidth]{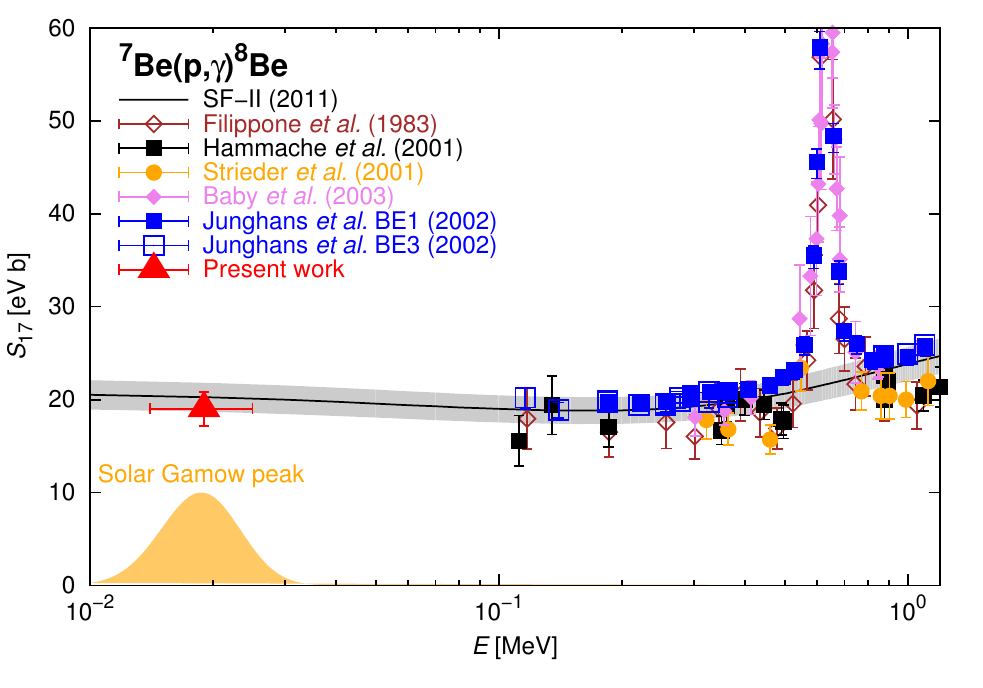}
\caption{$S$-factor for the $^7$Be($p,\gamma$)$^8$B reaction from experiments \cite{Filippone83-PRC,Hammache01-PRL,Strieder01-NPA,Baby03-PRL,Junghans02-PRL} and the SF-II evaluation \cite{Adelberger11-RMP}. The grey shaded area corresponds to the 1$\sigma$ error band of the SF-II evaluation \cite{Adelberger11-RMP}. The present neutrino-based result for $S_{17}$ is added to the plot. The solar Gamow peak is also shown.}
\label{fig:s17factor}
\end{figure}

Experimental data for $S_{17}(E)$ are available at $E > 0.1$ MeV \cite{Filippone83-PRC, Hammache98-PRL, Hammache01-PRL, Strieder01-NPA, Baby03-PRL,Baby03-PRC,Baby04-PRC,Junghans02-PRL, Junghans03-PRC, Junghans10-PRC} (see Fig.~\ref{fig:s17factor}). The energy of the solar Gamow window, which lies at 0.019 MeV, was not reached by any of the experiments. Based on an extrapolation, the value of $S_{17}(0)$ was estimated to be
\begin{equation}
S_{17}(0)=20.8 \pm 0.7_{\text{expt}} \pm 1.4_{\text{theor}} \text{ eV b}
\end{equation}
in the ''Solar Fusion II'' workshop, hereafter called SF-II \cite{Adelberger11-RMP}. Recently, a consistent but more precise value of 21.3$\pm$0.7\,eV\,b has been obtained using halo effective field theory to fit the experimental data \cite{Zhang15-PLB}.

The lack of experimental data in the low-energy region can be addressed with the help of the latest experimentally measured solar neutrino fluxes, using the standard solar model to determine the nuclear parameters, here the $^7$Be($p,\gamma$)$^8$B $S$-factor.  A related idea has recently been explored for the case of the $^3$He($\alpha$,$\gamma$)$^7$Be reaction \cite{Takacs15-PRD}. 

The contribution to the pp-III chain controlled by the $^7$Be($p,\gamma$)$^8$B reaction on the total $^4$He and energy production is very small, but a branch of solar neutrinos emerges from the decay of $^8$B \cite{SuperK98-PRL,SNO01-PRL,SNO02-PRL}. The flux of these neutrinos  is well-measured: The Sudbury Neutrino Observatory (SNO) reports a $^8$B solar neutrino flux of
\begin{equation*}
\phi^{\text{exp}}_{\text{B}} = 5.25 \pm 0.16 \text{ (stat)} ^{+0.11}_{-0.13} \text{ (sys)} \times 10^6 {\rm cm}^{-2}{\rm s}^{-1}
\end{equation*}
taking into account the loss in the amount of electron neutrinos due to the mixing among the neutrino flavours \cite{SNO13-PRC_all3phases}. This is equivalent to 3.9\% precision (systematical and statistical uncertainties combined in quadrature) and consistent with the determination made by Super-Kamiokande \cite{SuperK13-NPBPS}. 

In addition, the  Borexino collaboration very recently reported an improved result on the flux of $\bes$  neutrinos  \cite{Borexino17-arXiv}:  
\begin{equation*}
\phi^{\text{exp}}_{\text{Be}} = 4.99 \pm 0.13 {\rm (stat)} ^{+0.07}_{-0.10} {\rm (syst)} \times 10^9 {\rm cm}^{-2}{\rm s}^{-1}
\end{equation*}
again assuming neutrino flavour mixing, which corresponds to a precision of 3.1\%.

\section{Approach}
As in Ref. \cite{Takacs15-PRD}, the Standard Solar Model (SSM) by Bahcall et al. \cite{Bahcall89-Book, Bahcall06-APJSS}, as updated by the Barcelona group \cite{Serenelli13-PRD, Vinyoles17-ApJ}, is used here, and no new solar model calculations are performed.  The SSM uses a number of input parameters $(\beta_j)$, including the solar age ($A$), luminosity ($L$), opacity ($O$), diffusion rate ($D$), the key astrophysical \textit{S}-factors, and the zero-age abundance of chemical elements ($x_i$). The sensitivity of the neutrino flux $\phi_{i}$ for an arbitrary variation in a parameter $\beta_{j}$ can be expressed by the logarithmic partial derivatives $\alpha(i,j)$ given by
\begin{equation} \label{eq:Def_alpha}
\alpha(i,j)=\frac{\partial ln[\phi_{i}/\phi^{\rm SSM}_{i}]}{\partial ln[\beta_{j}/\beta^{\rm SSM}_{j}]}
\end{equation}
where $\phi^{\rm SSM}_{i}$ and $\beta^{\rm SSM}_{j}$ represent the SSM theoretical values. The SSM best values for the fluxes have been recently updated by the B16 SSM calculation by Vinyoles {\it et al.} \cite{Vinyoles17-ApJ}. For the present purposes, the partial derivatives $\alpha(i,j)$ are taken from the previous SSM by Serenelli {\it et al.} \cite{Serenelli13-PRD}. It is stated in B16 \cite{Vinyoles17-ApJ} that their partial derivatives are consistent with Serenelli {\it et al.} \cite{Serenelli13-PRD}. Using all this information, small changes in the neutrino fluxes may be approximated by a power law: 
\begin{equation} \label{eq:PowerLaw}
\frac{\phi_{i}}{\phi^{\rm SSM}_{i}}=\prod \limits_{j} \limits^{N} \Big(\frac{\beta_{j}}{\beta^{\rm SSM}_{j}}\Big) ^{\alpha(i,j)}
\end{equation}

It has to be noted that the flux predictions strongly depend on another output parameter of the SSM, namely the core temperature ($T_c$) of the Sun \cite{Bahcall96-PRD}:
\begin{equation}
\phi(^7{\rm Be}) \propto T_c^{11}
\end{equation}
\begin{equation}
\phi(^8{\rm B}) \propto T_c^{25}
\end{equation}
As it can be seen, the dependence is significantly stronger for $^8$B, than for $^7$Be.

One way that has been proposed in the literature \cite{Haxton08-APJ} to decrease this strong ($\phi_i \propto {T_c}^{a_i}$) dependence on the $T_c$ is to take Eq.~\ref{eq:PowerLaw} for two different branches of solar neutrinos ($i=$ 1, 2) and form a weighted ratio of their temperature exponents: $\phi_1 /\phi_2^{b}$, where the exponent $b$ can be approximated as $b \approx a_1/a_2$. However, one has to take into account the correlation between the neutrino fluxes $\phi_1$ and $\phi_2$ \cite{Serenelli13-PRD}. Therefore, $b$ is to be determined by minimizing the dependence of the flux ratio on the net logarithmic derivatives instead \cite{Haxton08-APJ}. 

The parameter of interest in the present work is the $S$-factor of  the $^7$Be($p,\gamma$)$^8$B reaction, here denoted as $S_{17}$.  By using the corresponding logarithmic partial derivatives from Serenelli {\it et al.} (Tables I and II in  \cite{Serenelli13-PRD}), one finds the following relation:

\begin{align}
 \label{eq:serenelli-eq}
 \frac{\phi^{\text{exp}}_{\text{B}}}{\phi^{\text{SSM}}_{\text{B}}} & \bigg/ \left[\frac{\phi^{\text{exp}}_{\text{Be}}}{\phi^{\text{SSM}}_{\text{Be}}}\right]^{2.151} = 
 \nonumber\\ 
 & \left[L^{-0.471} O^{0.009} A^{-0.289} D^{-0.004}\right] \nonumber\\ 
 & \times \left[S^{-0.449}_{11}   S^{0.515}_{33} S^{-1.028}_{34} S^{1.000}_{17} S^{-1.000}_{e7} S^{0.009}_{114}\right] \nonumber\\
 &  \times \left[x^{0.023}_{\text{C}}  x^{0.005}_{\text{N}} x^{0.006}_{\text{O}} x^{-0.009}_{\text{Ne}} x^{-0.016}_{\text{Mg}} x^{-0.032}_{\text{Si}} x^{-0.023}_{\text{S}} x^{-0.006}_{\text{Ar}} x^{0.057}_{\text{Fe}}\right] 
\end{align}

\begin{equation}\label{eq:final}
S_{17}=S_{34}^{1.028}S_{e7}\left[\frac{S_{11}^{0.449}}{S_{33}^{0.516}}\right] F_{\rm SSM}   \frac{\phi^{\text{exp}}_{\text{B}}}{\phi^{\text{SSM}}_{\text{B}}} \bigg/ \left[\frac{\phi^{\text{exp}}_{\text{Be}}}{\phi^{\text{SSM}}_{\text{Be}}}\right]^{2.151} 
\end{equation}

\begin{align}
F_{\rm SSM}= & \left[L^{0.471} O^{-0.009} A^{0.289} D^{0.004}\right]   \nonumber\\
		\times &  \left[x^{-0.023}_{\text{C}} x^{-0.005}_{\text{N}} x^{-0.006}_{\text{O}} x^{0.009}_{\text{Ne}} x^{0.016}_{\text{Mg}} x^{0.032}_{\text{Si}} x^{0.023}_{\text{S}} x^{0.006}_{\text{Ar}} x^{-0.057}_{\text{Fe}}\right]
\end{align}

In the equation, $S_{11}$, $S_{33}$, $S_{34}$, $S_{17}$, $S_{e7}$, and $S_{114}$   are the astrophysical $S$-factors of the $^1$H(p,e$^+\nu_e$)$^2$H, $^3$He($^3$He,2p)$^4$He, $^3$He($\alpha,\gamma$)$^7$Be, $^7$Be($p,\gamma$)$^8$B, $^7$Be($e,\nu_e$)$^7$Li, and $^{14}$N($p,\gamma$)$^{15}$O   reaction,  respectively, normalised to their SF-II recommended value. $x_i$ is the solar core abundance of element $i$.

An interesting fact is that the $\phi_{\text{B}}/\phi^{2.151}_{\text{Be}}$ flux ratio is almost independent of the choice of the solar elemental composition (GS98 \cite{Grevesse98-SSR} or AGSS09  \cite{Asplund09-ARAA}), because the competition between electron capture and proton capture on $^7$Be is independent of metallicity (see Table~\ref{tab:fluxes}). Thus, the left hand side of the  Eq.~\ref{eq:serenelli-eq} is unaffected by the solar metallicity problem. This is reflected in very low exponents $\ll$1 for all the elemental abundances $x_i$ in Eq.~\ref{eq:serenelli-eq}.

Solving for the astrophysical $S$-factor $S_{17}$, one obtains Eq.~\ref{eq:final} where the remaining contributions from non-nuclear parameters were collected into  $F_{\rm SSM}$ and the negligible  dependence on $S_{114}$ was disregarded.

%============================================================================

%============================================================================
\section{Nuclear physics based determination of $S_{34}$}
\begin{figure}
\centering
\includegraphics[width=\columnwidth]{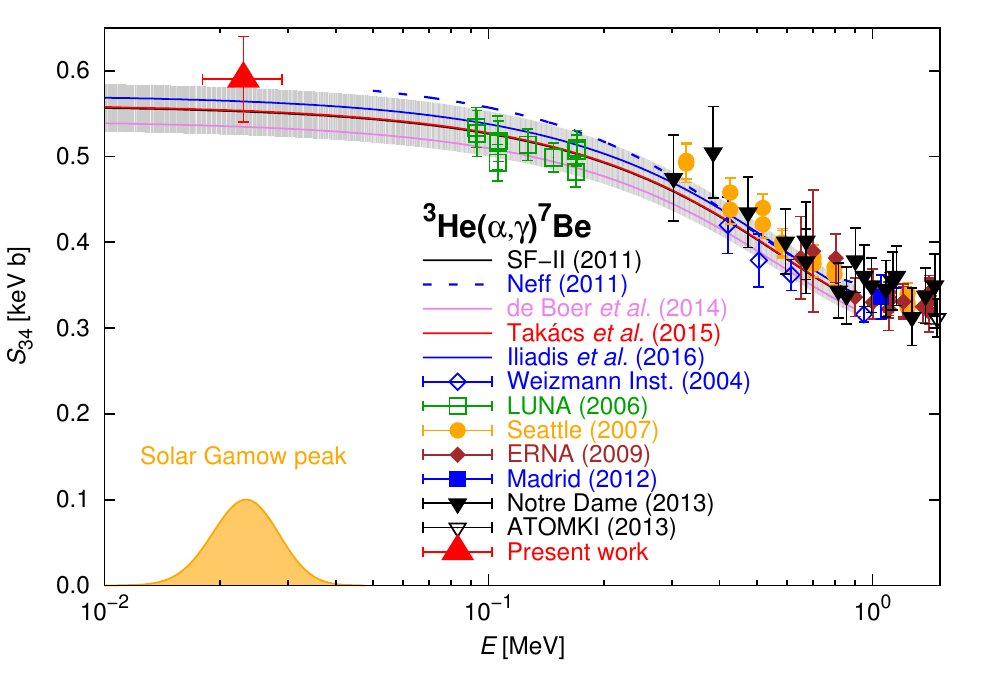}
\caption{Cross section of the $^3$He($\alpha,\gamma$)$^7$Be reaction, parametrized as the astrophysical $S$-factor, from experiment \cite{NaraSingh04-PRL,Bemmerer06-PRL,Confortola07-PRC,Brown07-PRC,diLeva09-PRL,Carmona12-PRC,Kontos13-PRC,Bordeanu13-NPA},  theory \cite{Neff11-PRL}, or evaluations \cite{Adelberger11-RMP,deBoer14-PRC,Takacs15-PRD,Iliadis16-ApJ}. The grey shaded area corresponds to the 1$\sigma$ error band of the SF-II evaluation. The present neutrino-based result for $S_{34}$ is added to the plot. The solar Gamow peak is also shown.}
\label{fig:s34factor}
\end{figure}

Of the nuclear input in Eq.~\ref{eq:final}, the $S$-factors of the $^3$He($\alpha,\gamma$)$^7$Be and $^3$He($^3$He,$2p$)$^4$He reactions come with the largest uncertainties. However, due to the difference in their exponents, the nuclear error budget is dominated by $S_{34}$ alone. Originally this uncertainty was estimated to be $5.4\%$ in SF-II \cite{Adelberger11-RMP}. Recently, de Boer \textit{et al.} \cite{deBoer14-PRC} and Iliadis \textit{et al.}  \cite{Iliadis16-ApJ} re-evaluated $S_{34}$ based on a global R-matrix fit and a global  Bayesian analysis, respectively, resulting in slightly different, but still consistent $S_{34}(0)$ values and lower uncertainties (see Fig.~\ref{fig:s34factor}). However, the estimated solar neutrino fluxes from the B16 SSM calculation of \cite{Vinyoles17-ApJ} are still based on SF-II. 

Therefore, the SF-II approach is retained here, extending the original  compilation with the most recent experimental data from the University of Notre Dame \cite{Kontos13-PRC}. The previous result from solar neutrino based determination of the $^3$He($\alpha,\gamma$)$^7$Be $S$-factor \cite{Takacs15-PRD} is set aside here, because it would introduce a circular logic into the evaluation.
Now, by following the same fitting approach as in SF-II \cite{Adelberger11-RMP}, one finds, based purely on nuclear physics data (and excluding for now neutrino data):
\begin{equation}
S_{34}(0)=0.561\pm0.014_{\textnormal{expt}}\pm0.02_{\textnormal{theor}} \text{ keV b}
\end{equation}
where the experimental error includes both statistical and systematic uncertainties related to the experiments, while the theoretical error arises from the fact the experimental results had to be extrapolated to the solar Gamow peak with the help of a theoretical model \cite{Adelberger11-RMP}. 

The present experimental error of $\pm0.014$\,keV\,b is slightly lower than the SF-II value of $\pm0.016$\,keV\,b \cite{Adelberger11-RMP} due to the addition of the post-SF-II Notre Dame experimental data \cite{Kontos13-PRC}. These data have 8\% systematic error and thus only a minor effect on the central $S_{34}(0)$ value. However, their inclusion somewhat improves the overall confidence in the fit. $\chi^2$/dof changes from 4.6/3 to 5.6/4, and $P(\chi^2,{\rm dof})$ from 0.07 to 0.10. Thus the inflation factor \cite{Adelberger11-RMP} applied to the uncertainty of the average value reduces from 1.73 to 1.52, and the final $\Delta_{\textnormal{expt}}S_{34}(0)$ value from 0.016 to 0.014 keV\,b.
The total uncertainty on  $S_{34}$ is now $4.3\%$, following the SF-II approach and simply adding the post-SF-II Notre Dame \cite{Kontos13-PRC} experimental data.

Both the $S_{34}(0)$ and its error bar are consistent with the estimates provided by de Boer \textit{et al.} \cite{deBoer14-PRC} and Iliadis \textit{et al.} \cite{Iliadis16-ApJ}: They find 0.542 $\times$(1 $\pm4.2\%$) and 0.572 $\times$(1 $\pm3.1\%$), respectively.  Adding the 4.3\% uncertainty of the present $S_{34}$ value in quadrature to the uncertainty of $S_{11}$, $S_{33}$ and $S_{e7}$, the total nuclear contribution to the error budget amounts to $5.6\%$.

\begin{table}[h]

\caption{Predicted solar neutrino fluxes from the SSM for two different elemental abundances, taken from \cite{Vinyoles17-ApJ}. The flux ratio (Eq.~\ref{eq:serenelli-eq}) is almost independent of the assumption on metallicities.}
\smallskip

\begin{center}

\begin{tabular}{lccc}
\hline\noalign{\smallskip}
Elemental comp.  & $\phi^{\text{SSM}}_{\text{B}}$ & $\phi^{\text{SSM}}_{\text{Be}}$ & $[\phi^{\text{exp}}_{\text{B}}/\phi^{\text{SSM}}_{\text{B}}]/[\phi^{\text{exp}}_{\text{Be}}/\phi^{\text{SSM}}_{\text{Be}}]^{2.151}$ \\
\noalign{\smallskip}
\hline
\hline
\noalign{\smallskip}
GS98 \cite{Grevesse98-SSR} & 5.46 & 4.93 & 	0.937\\
AGSS09 \cite{Asplund09-ARAA} & 4.50 & 4.50 & 0.934\\
\noalign{\smallskip}
\hline
\end{tabular}

\end{center}
\label{tab:fluxes}
\end{table}

%============================================================================

\section{Solar neutrino based determination of $S_{17}$}

As the next step, the nuclear physics based $S_{34}$ value is used to determine $S_{17}$ from Eq.~\ref{eq:final}. Here, the dominant error bar is given by the $^7$Be neutrino flux, and its exponent of 2.151. This considerably inflates the $3.1\%$ uncertainty of  $\Phi^{\rm exp}_{\rm Be}$. Therefore, this part of the equation will contribute $7.7\%$ towards the total error.

Summing up the contributions of all three main sources of error, one finds that the method followed here provides an estimate on  $S_{17}$ with total uncertainty of $9.6\%$. The S-factor of the $^7$Be($p,\gamma$)$^8$B reaction is then redetermined in the solar Gamow peak by accepting the previous  SF-II recommendation \cite{Adelberger11-RMP}, and rescaling according to Eq.~\ref{eq:final}:
\begin{equation}
   S_{17}(19^{+6}_{-5})=19.0\pm1.8 \text{ eV b}
\end{equation}
In order to facilitate a comparison with other work, this value can be converted to the zero-energy S-factor by using the SF-II fit curve:
\begin{equation}
   S_{17}(0)=19.5\pm1.9 \text{ eV b}
\end{equation}
 
\section{Solar neutrino based determination of $S_{34}$}

Additionally, the previous estimate on the S-factor of the $^3$He($\alpha,\gamma$)$^7$Be is updated here, using the new experimental flux of $^7$Be neutrinos  \cite{Borexino17-arXiv} and the B16 SSM calculated $^7$Be neutrino flux,  otherwise following the approach of Ref. \cite{Takacs15-PRD}:   
\begin{equation}
   S_{34}(23^{+6}_{-5})=0.590\pm 0.050 \text{ keV b}
\end{equation}
Again using the SF-II fit curve, this value can be converted to the zero-energy S-factor:
\begin{equation}
   S_{34}(0)=0.598\pm 0.051 \text{ keV b}
\end{equation}

\section{Discussion, summary, and outlook}

The two new, neutrino-based S-factor determinations are included in Figures \ref{fig:s17factor} and \ref{fig:s34factor}. 

For the case of $S_{17}$, the new, neutrino-based value is lower than, but still consistent with, the nuclear-physics based extrapolation given in SF-II (Fig. \ref{fig:s17factor}). The present error bar is slightly higher, however it has to be stressed that the present result uses a completely different approach, without any nuclear experiment on the $^7$Be($p,\gamma$)$^8$B  reaction. So this independent result provides a valuable confirmation of the correctness of the nuclear physics experiments and their extrapolations.

For the case of $S_{34}$, the new, neutrino-based value is higher than, but still consistent with, all of the previous nuclear-physics based evaluations, namely SF-II \cite{Adelberger11-RMP}, SF-II (updated here), de Boer \cite{deBoer14-PRC}, and Iliadis \citep{Iliadis16-ApJ} (Fig. \ref{fig:s34factor}). Again, the present result furnishes a completely independent confirmation of the correctness of the previous, nuclear-based evaluations \cite{Adelberger11-RMP, deBoer14-PRC, Iliadis16-ApJ}.

Summarizing, the very recent, highly precise $^7$Be neutrino flux data from Borexino  \cite{Borexino17-arXiv} has been used in conjunction with the $^8$B neutrino flux from SNO and the standard solar model to determine, based on solar neutrino fluxes, the astrophysical $S$-factor of the $^7$Be($p,\gamma$)$^8$B reaction. In addition, a previous neutrino-based determination of the $^3$He($\alpha,\gamma$)$^7$Be $S$-factor \cite{Takacs15-PRD} has been updated here as well. The new data independently confirm the most recent nuclear physics based determinations of these two astrophysical $S$-factors.

It is now up to nuclear physics experimenters and theoreticians to further improve the precision of these two astrophysical $S$-factors, to enable a possible future application of the neutrino flux data to study the solar metallicity problem \cite{Haxton08-APJ}.

Financial support by the Helmholtz Association (HGF) through the Nuclear Astrophysics Virtual Institute (NAVI, HGF VH-VI-417) and by DFG (BE 4100/4-1 and ZU 123/21-1) is gratefully acknowledged.

\bibliographystyle{elsarticle-num_etal}
\bibliography{Danielsbib,/Users/Marcell/Documents/Science/Literature/MTakacs_library}

%============================================================================

%============================================================================

\end{document}